\documentclass[aps,prd,superscriptaddress,twocolumn,showpacs]{revtex4}
\usepackage{amsfonts,amssymb,amscd,amsmath}
\usepackage{graphicx}
\usepackage{epsfig}
\usepackage{latexsym}
\usepackage{amssymb}
\usepackage{subfigure}
\usepackage{mathrsfs}
\usepackage{extarrows}
\usepackage{dsfont}

\begin{document}

\def\abs#1{ \left| #1 \right| }
\def\lg#1{ | #1 \rangle }
\def\rg#1{ \langle #1 | }
\def\lr#1{ \langle #1 \rangle }
\def\lrg#1#2#3{ \langle #1 | #2 | #3 \rangle }

\newcommand{\hf}{\frac{1}{2}}
\newcommand{\bra}[1]{\left\langle #1 \right\vert}
\newcommand{\ket}[1]{\left\vert #1 \right\rangle}
\newcommand{\bx}{\begin{matrix}}
\newcommand{\ex}{\end{matrix}}
\newcommand{\be}{\begin{eqnarray}}
\newcommand{\ee}{\end{eqnarray}}
\newcommand{\nn}{\nonumber \\}
\newcommand{\no}{\nonumber}
\newcommand{\de}{\delta}
\newcommand{\Del}{\Delta}
\newcommand{\lt}{\left\{}
\newcommand{\rt}{\right\}}
\newcommand{\lx}{\left(}
\newcommand{\rx}{\right)}
\newcommand{\lz}{\left[}
\newcommand{\rz}{\right]}
\newcommand{\pu}{\partial_{\mu}}
\newcommand{\pv}{\partial_{\nu}}
\newcommand{\au}{A_{\mu}}
\newcommand{\av}{A_{\nu}}
\newcommand{\p}{\partial}
\newcommand{\ts}{\times}
\newcommand{\ld}{\lambda}
\newcommand{\al}{\alpha}
\newcommand{\bt}{\beta}
\newcommand{\ga}{\gamma}
\newcommand{\si}{\sigma}
\newcommand{\ro}{\rho}
\newcommand{\tu}{\tau}
\newcommand{\z}{\mathrm}
\newcommand{\dg}{\dagger}
\newcommand{\Og}{\Omega}
\newcommand{\Ld}{\Lambda}
\newcommand{\m}{\mathcal}
\newcommand{\bb}{\mathbf}

\title{Bounds on Quantum Multiple-Parameter Estimation with Gaussian State}

\author{Yang Gao}\email{gaoyangchang@gmail.com}
\affiliation{\it Department of Physics, Xinyang Normal University,
Xinyang, Henan 464000, China}

\author{Hwang Lee}
\affiliation{Hearne Institute for Theoretical Physics and
Department of Physics and Astronomy, \\
Louisiana State University, Baton Rouge, LA 70803, USA }

\begin{abstract}{}
We investigate the quantum Cramer-Rao bounds on the joint
multiple-parameter estimation with the Gaussian state as a probe. We
derive the explicit right logarithmic derivative and symmetric
logarithmic derivative operators in such a situation. We compute the
corresponding quantum Fisher information matrices, and find that
they can be fully expressed in terms of the mean displacement and
covariance matrix of the Gaussian state. Finally, we give some
examples to show the utility of our analytical results.
\end{abstract}

\pacs{03.65.Ta, 03.67.-a, 06.20.Dk, 42.50.St}

\maketitle

\section{Introduction}

Employing quantum resources to improve the sensitivity in the
estimation of relevant physical parameters is of great importance in
metrology and sensing \cite{glm}. Much of the work has focused on
the estimation of a single parameter, both theoretically and
experimentally \cite{glm,excep}. However, there are many situations
where the joint estimation of multiple parameters becomes necessary,
e.g., the vector phase estimation in recent field of phase imaging
and microscopy \cite{image}. The developed multi-port devices and
multi-qubit manipulation also demand the investigation of
multi-parameter sensitivity from theoretical viewpoint.

The joint estimation of multiple parameters is an example of the
general problem of quantum estimation theory \cite{qet}. A typical
parameter estimation consists in sending a probe in a suitable
initial state through some parameter-dependent physical process and
measuring the final state of the probe, estimating then from this
measurement the values of the parameters. Consider a family of
quantum state $\rho_{\theta}$ which depend on a set of $d$ different
parameters $\theta=(\theta_1,...,\theta_d)^T$. The aim of quantum
estimation theory is to infer the values of $\theta$ from the
outcomes of a generalized measurement $M$ (characterized by POVM
$\{M_\xi\}$, $\xi=(\xi_1,\xi_2,...)^T$, $M_\xi \ge 0$, $\int d\xi
M_{\xi}=\mathds 1$). Let $\Theta(\xi)$ be the estimator of $\theta$
constructed from the outcome $\xi$. To quantify the sensitivity of
this estimation, a local covariance matrix is defined as
$V_\theta(M) = \int d\xi (\Theta(\xi)-\theta)(\Theta(\xi)-\theta)^T
p({\xi}|{\theta})$, where $p(\xi|\theta)={\rm Tr}[\rho_{\theta}
M_{\xi}]$ is the conditional probability distribution of obtaining a
certain outcome $\xi$ given $\theta$. From now on we assume a
particular point $\theta$ and consistently drop the dependency on
$\theta$.

In order to present the lower bounds for $V(M)$, one can define the
so-called right logarithmic derivative (RLD) and symmetric
logarithmic derivative (SLD) operators for each of the parameters
involved \cite{qet}, respectively as \be \p_k \rho
&=& \rho \m L_k \quad ({\rm RLD}), \label{rlde} \\
\p_k \rho &=& \hf (\rho \mathscr L_k+\mathscr L_k \rho) \quad ({\rm
SLD}), \ee  where $\p_k \equiv \p/\p\theta_k$. Then one can define
two matrices \be \mathcal F_{ij} &=& {\rm Tr}[\rho\mathcal L_i
\mathcal L_j^\dg] = \m F_{ji}^*, \label{rqfi} \\ \mathscr F_{ij} &=&
\hf {\rm Tr}[\rho(\mathscr L_i \mathscr L_j+\mathscr L_i \mathscr
L_j)] = \mathscr F_{ji}, \ee which are called the RLD and SLD
quantum Fisher information (QFI) matrices, respectively. The SLD QFI
matrix can be computed from Uhlmann's quantum fidelity between two
outgoing final states corresponding to two different sets of
parameters \cite{uncert}. By defining a positive definite matrix
$G$, two diffferent Cramer-Rao bounds hold \cite{qet}, \be {\rm Tr}
[G V(M)] & \ge & \frac{1}{\nu} \lx {\rm
Tr} [G \mathcal F_R^{-1}]+{\rm Tr} [|G \mathcal F_I^{-1}|] \rx, \\
{\rm Tr} [GV(M)] & \ge & \frac{1}{\nu} {\rm Tr} [G\mathscr F^{-1}],
\label{sldb} \ee where $|A| \equiv \sqrt{AA^\dg}$ and $\nu$ is the
number of the measurements performed. If we choose $G=\mathds 1$, we
obtain the two bounds on the sum of the variances of the parameters
involved, \be \nu \sum_{k=1}^d \delta^2 \theta_k &\ge & B_R \equiv
{\rm Tr}[\m
F_R^{-1}]+{\rm Tr}[|\m F_I^{-1}|], \label{rbound} \\
\nu \sum_{k=1}^d \delta^2 \theta_k &\ge & B_S \equiv {\rm Tr}
[\mathscr F^{-1}]. \label{sbound} \ee

In general, neither the RLD bound nor the SLD bound is attainable
\cite{am1}. Here the fundamental non-commutativity of quantum theory
forbids simultaneously obtaining the optimal estimations of all
parameters, and optimizing the measurement for one parameter will
usually disturb the measurement precision on the others. At the same
time, the optimal estimator for the RLD bound might not correspond
to a POVM. On the other hand, even if the optimal measurements for
both parameters do not commute, it is still possible to attain both
bounds by devising a single simultaneous measurement.

Most of the work on quantum parameter estimation are devoted to the
SLD bound \cite{am1,am,am2,dampphase} (notable contributions are
Refs. \cite{multiple,pomo}). For single-parameter estimation
problems, it is known that the SLD QFI is always smaller than the
RLD QFI, and thus gives a tighter bound for parameter sensitivity
\cite{qet}. Moreover, the single-parametric SLD QFI is
asymptotically attainable for large $N$. For multi-parametric cases,
recent progress in the theory of local asymptotic normality for
quantum states suggests that Eq. (\ref{sldb}) is asymptotically
attainable if and only if \be {\rm Tr} [\rho [\mathscr L_i, \mathscr
L_j]]=0. \label{asym} \ee When this condition can not be fulfilled,
the RLD bound could be tighter, and therefore becomes more
important.

The analytical expressions for the RLD and SLD operators and the QFI
matrices always pose formidable challenges, except for very
particular situations when the density state can be simply put in
the diagonal form \cite{excep,dampphase}. For the Gaussian state,
the author of Ref. \cite{am2} has obtained the explicit forms of the
SLD operator and the corresponding QFI matrix in terms of the mean
displacement and covariance matrix. In this paper, we generalized it
to the RLD case and derive the relevant results.

We organize this paper as follows. Section II reviews some basic
notations for the Gaussian state. Then in section III we derive the
expressions for the RLD and SLD operators and the QFI matrices for
multi-parameter estimation. In section IV we give some examples to
show the utility of our obtained results. Finally, we end with a
short summary.

\section{Notations for Gaussian state}

Consider $n$ bosonic modes with annihilation operators $a_i$
satisfying the commutation relations $[a_i,a_j^\dg]=\delta_{ij}$,
all other commutators being zero. Arrange all operators into a
vector $a^\mu=(a_1,a_1^\dg,a_2,a_2^\dg,\dots)^T$. From now on
Einstein's summation convention will be used throughout this paper.
The commutation relations are expressed as
$[a^\mu,a^\nu]=\Omega^{\mu\nu}$ and $a^{\mu \dg}= X^\mu_\nu a^\nu
\equiv  X^{\mu\nu} a^\nu  $, where $\Omega=\bigoplus_{k=1}^n \imath
\sigma_y$ and $X=\bigoplus_{k=1}^n \sigma_x$, in terms of the
$x,y$-component Pauli matrices $\sigma_{x,y}$. Note that
$\Omega^T=-\Omega$ and $\Omega^2=-\mathds 1$. Let us introduce the
convention $\Omega_{\mu\nu}=\Omega^{\mu\nu}$, so that
$\Omega^{\alpha\beta}\Omega_{\bt \ga}=-\delta^\al_\ga$, to lower
(rise) the index of $a^\mu$ ($a_\mu$) through
$a_\mu=\Omega_{\mu\nu}a^\nu$ and $a^\mu=\Omega^{\nu\mu}a_\nu$.
Equivalently, the position and momentum operators are arranged in
the vector $x^\mu=(q_1,p_1,q_2,p_2,\dots)^T$, related to $a^\mu$ by
a unitary transformation $x^\mu=H_\nu^\mu a^\nu \equiv
H^{\mu\nu}a^\nu$, where
\be H=\bigoplus_{k=1}^n \frac{1}{\sqrt{2}} \bigg( \bx 1 & 1 \\
-\imath & \imath \ex \bigg), \quad H^{\dg}H=1. \ee

The Gaussian state is defined as a state with Gaussian
characteristic function \be \chi(z)=\z{Tr}[\rho e^{-z_\mu a^\mu}]=
\exp\lx \hf\Sigma^{\mu\nu} z_\mu z_\nu- \lambda^\mu z_\mu \rx, \ee
which is fully described by the mean displacement and covariance
matrix \be \lambda^\mu &=& \z{Tr}[\rho a^\mu] \\
\Sigma^{\mu\nu} &=& \hf \z{Tr}[\rho(\tilde{a}^\mu \tilde{a}^\nu
 +\tilde{a}^\nu \tilde{a}^\mu)] \ee in terms of the centered operator
$\tilde{a}^\mu = a^\mu-\lambda^\mu$. Here
$z^\mu=(z_1,z_1^*,z_2,z_2^*,\dots)^T$, $z^{\mu *}=X^\mu_\nu z^\nu$,
and $z_\mu=\Omega_{\mu\nu} z^\nu$.

With the above notations, the following identities will be
frequently used, \be
X^\mu_\nu \tilde{a}^\nu &=&\tilde{a}^{\mu \dg} \\
X^\mu_\nu \ld^\nu &=&\ld^{\mu *} \\
\z{Tr}[\rho\tilde{a}^\mu \tilde{a}^\nu] &=& \Sigma_+^{\mu\nu}\\
X^\mu_\gamma X^\nu_\delta
\Sigma^{\gamma\delta}&=&\Sigma^{\mu\nu*} \\
X^\mu_\gamma X^\nu_\delta
\Omega^{\gamma\delta}&=&-\Omega^{\mu\nu} \\
\Sigma^{\mu\nu}&=&\Sigma^{\nu\mu} \\
\Omega^{\mu\nu}&=&-\Omega^{\nu\mu} \\
\Sigma_+^{\mu\nu}&=&\Sigma_-^{\nu\mu}, \ee where $\Sigma_\pm \equiv
\Sigma \pm \Omega/2$.

\section{Derivation of the RLD and SLD operators}

In this section, we derive the RLD and SLD operators for the
Gaussian state. For simplicity, we first consider the RLD case. It
is known \cite{book} that a displaced gaussian state is related to a
centered gaussian state by $\rho_\ld = D(\ld) \rho_0 D^\dg(\ld)$,
where the unitary matrix $D(\ld)\equiv e^{\ld^\mu a_\mu}$ and $D
a^\mu D^\dg=\tilde{a}^\mu$. The right hand side of Eq. (\ref{rlde})
can be written as \be {\p_k \rho_\ld}=\p_k' \rho_\ld + D (\p_k
\rho_0) D^\dg, \label{split} \ee where the symbol $\p_k'$ in the
first term means only taking the derivative of $\ld^\mu$ with
respect to $\theta_k$, and the second term corresponds to the RLD
for the centered state.

For the first term, since the RLD of a Gaussian state with respect
to $\theta_k$ of $\ld^\mu$ involves the data linearly, we try the
linear form $\m L_k'=\m B^{(k)}_{\mu}a^\mu+\m C_k'$. Then we convert
Eq. (\ref{rlde}) into c-number equation through the characteristic
function and the operator rule \cite{book}, $\rho a^\mu \to \lx
-\p^\mu+ z^\mu /2 \rx \chi$, where $\p^\mu \equiv {\p/\p z_\mu}$.
This leads to \be \rho a^\mu &\to & \lx -\Sigma^{\mu
\nu}z_\nu+\ld^\mu+\hf \Omega^{\nu\mu}z_\nu \rx \chi \nn &\equiv &
(-\Sigma_-^{\nu\mu} z_\nu + \ld^\mu)\chi \ee and Eq. (\ref{rlde})
implies \be \Sigma_-^{\nu\mu}\m B^{(k)}_\mu &=& \p_k \ld^\nu, \\ \m
B^{(k)}_\mu \ld^\mu + \m C_k' &=& 0. \ee

Next we consider $\p_k \rho_0$ in the second term of Eq.
(\ref{split}). Since the RLD of a centered Gaussian state with
respect to $\theta_k$ of $\Sigma^{\mu\nu}$ involves the data
quadratically, we try the quadratic form $\m L_k''=\m
A^{(k)}_{\mu\nu}a^\mu a^\nu+\m C_k''$ with
$A^{(k)}_{\mu\nu}=A^{(k)}_{\nu\mu}$. We also convert Eq.
(\ref{rlde}) into c-number equation via the operator rule, \be
\rho a^\mu a^\nu & \to & \bigg[ \p^\mu \p^\nu + \hf
\Og^{\mu\nu}+{1\over 4} \Og^{\al \mu}\Og^{\bt \nu}z_\al z_\bt \nn
&& -\hf \lx \Og^{\al\nu} \p^\mu+\Og^{\al\mu} \p^\nu \rx z_\al
\bigg] \chi \nn &=& \lx \Sigma^{\mu\nu}+\hf \Og^{\mu\nu}\rx \chi
+\bigg [ \Sigma^{\al\mu}\Sigma^{\bt\nu}+{1\over
4}\Og^{\al\mu}\Og^{\bt\nu} \nn &&
-\hf(\Sigma^{\al\mu}\Og^{\bt\nu}+ \Og^{\al\mu}\Sigma^{\bt\nu})
\bigg ] z_\al z_\bt \chi \nn &=& \Sigma_+^{\mu\nu}\chi+ \mathsf
M^{\al\bt,\mu\nu} z_\al z_\bt \chi, \ee where $\mathsf M \equiv
\Sigma_- \otimes \Sigma_-$. So Eq. (\ref{rlde}) implies \be
\mathsf M^{\al\bt,\mu\nu} \m A^{(k)}_{\mu\nu}&=&\hf\p_k
\Sigma^{\al\bt}, \\ \m A^{(k)}_{\mu\nu} \Sigma_+^{\mu\nu}+\m C_k''
&=& 0.\ee Combining the above results, the RLD takes \be \m L_k
&=& \widetilde{\m L}_k'' +\m L_k' = D \m L_k'' D^ \dg +\m L_k' \nn
&=& \m A^{(k)}_{\mu\nu}(\tilde{a}^\mu \tilde{a}^\nu -
\Sigma^{\mu\nu}) + \m B^{(k)}_{\mu}\tilde{a}^{\mu}, \label{rrld}
\ee where \be \m A^{(k)}_{\mu\nu} &=& \hf \mathsf
M^{-1}_{\mu\nu,\al\bt} \p_k\Sigma^{\al\bt}, \\ \m B^{(k)}_{\mu}
&=& (\Sigma_-)^{-1}_{\mu\nu} \p_k\lambda^\nu. \ee

Then, using the Wick theorem for the Gaussian state \footnote{If
$(x_1, бн, x_{2n})$ is a zero mean multivariate normal random
vector, then $E[x_1x_2\dots x_{2n}]=\sum\prod E[x_i x_j]$ and
$E[x_1x_2\dots x_{2n-1}]=0$, where the symbol $E[O]$ indicates
taking the expectation value of $O$, and the notation $\sum\prod$
means summing over all distinct ways of partitioning $x_1$, $x_2$,
$\dots$, $x_{2n}$ into pairs. This yields $(2n)!/(2^n n!)$ terms
in the sum. For $n=2$, it gives $
E[x_1x_2x_3x_4]=E[x_1x_2]E[x_3x_4]+E[x_1x_3]E[x_2x_4]+E[x_1x_4]E[x_2x_3]
$.}, we compute the RLD QFI matrix. Since the odd terms of the
cross products of $\widetilde{\m L}_k''$ and $\m L_k'$ vanish
identically, we only consider the terms involving the double
contributions of  $ \widetilde{\m L}_k''$ and $\m L_k'$,
respectively. Noticing \be\mathrm{Tr}[\rho \tilde{a}^\al \tilde
a^\bt \tilde a^\mu \tilde a^\nu] =
\Sigma_+^{\al\bt}\Sigma_+^{\mu\nu}+\Sigma_+^{\al\mu}
\Sigma_+^{\bt\nu}+\Sigma_+^{\al\nu}\Sigma_+^{\bt\mu},\ee we have
\be && {\rm Tr}[\rho \widetilde{\m L}_i'' \widetilde{\m
L}_j''^\dg] \nn && \xlongequal[]{(16,14)} \m A^{(i)}_{\al\bt}\m
A^{(j)*}_{\mu\nu}\lx X^\mu_\gamma
X^\nu_\delta\mathrm{Tr}[\rho\tilde a^\al\tilde a^\bt\tilde
a^\gamma\tilde a^\delta]- \Sigma^{\al\bt}\Sigma^{\mu\nu*}\rx \nn
&& \xlongequal[]{(17,32)} \m A^{(i)}_{\al\bt}\m A^{(j)*}_{\mu\nu}
X^\mu_\gamma X^\nu_\delta \lx \Sigma_+^{\al\gamma}
\Sigma_+^{\bt\delta}+\Sigma_+^{\al\delta}\Sigma_+^{\bt\gamma} \rx
\nn && \xlongequal[]{(21)} X^\mu_\gamma X^\nu_\delta \lx
\Sigma_-^{\gamma\al}
\Sigma_-^{\delta\bt}+\Sigma_-^{\delta\al}\Sigma_-^{\gamma\bt}\rx
\m A^{(i)}_{\al\bt}\m A^{(j)*}_{\mu\nu} \nn && \xlongequal[]{(27)}
{1\over 2} X^\mu_\gamma X^\nu_\delta \lx \p_i
\Sigma^{\gamma\delta}+\p_i \Sigma^{\delta\gamma} \rx \m
A^{(j)*}_{\mu\nu} \nn && \xlongequal[]{(17,19)} \lx \p_i
\Sigma^{\mu\nu*}\rx \m A^{(j)*}_{\mu\nu} \xlongequal[]{(3)} \lx
\p_j \Sigma^{\mu\nu}\rx \m A^{(i)}_{\mu\nu} \nn &&
\xlongequal[]{(30)} {1\over 2} \lx \p_j \Sigma^{\mu\nu} \rx
\mathsf M^{-1}_{\mu\nu,\al\bt} \p_i \Sigma^{\al\bt}, \ee where the
number over the equal sign means the corresponding equation has
been used. On the other hand, \be && {\rm Tr}[\rho \m L_i'{\m
L_j'}^\dg] \xlongequal[]{(14)} \m B^{(i)}_{\al}\m
B^{(j)*}_{\mu}X^\mu_\gamma\mathrm{Tr}[\rho\tilde a^\al\tilde
a^\gamma] \nn && \xlongequal[]{(16)} X^\mu_\gamma
\Sigma_+^{\al\gamma}\m B^{(i)}_{\al}\m B^{(j)*}_{\mu}
\xlongequal[]{(21)} X^\mu_\gamma \Sigma_-^{\gamma\al}\m
B^{(i)}_{\al}\m B^{(j)*}_{\mu} \nn && \xlongequal[]{(24)}
X^\mu_\gamma (\p_i \lambda^\gamma) \m B^{(j)*}_{\mu}
\xlongequal[]{(15)} (\p_i \lambda^{\mu*}) \m B^{(j)*}_{\mu} \nn &&
\xlongequal[]{(3)} (\p_j \lambda^{\mu}) \m B^{(i)}_{\mu}
\xlongequal[]{(31)} (\p_j \lambda^{\mu})
(\Sigma_-^{-1})_{\mu\nu}(\p_i \lambda^\nu). \ee The final
expression for the RLD QFI matrix is thus obtained \be \m F_{ij}=
\hf \mathsf M^{-1}_{\al\bt,\mu\nu} \p_j\Sigma^{\al\bt}
\p_i\Sigma^{\mu\nu}+(\Sigma_-)^{-1}_{\mu\nu}\p_j
\lambda^\mu\p_i\lambda^\nu. \label{rb} \ee According to Ref.
\cite{qet}, the RLD bound is attainable if there is a POVM
$\{M_{\hat{\theta}} \}$, so that for a particular vector
$\mathbf{Y}$, the following equation \be \sum_k (\mathbf{Y}^\dg \m
F^{-1})_k \m L_k M_{\hat{\theta}}=c
\sum_k(\hat{\theta}_k-\theta_k)M_{\hat{\theta}} \ee holds for some
complex number $c$, which may be a function of $\theta$. The
specific case for the RLD bound of estimating the parameters of a
coherent signal in thermal background has been obtain in Ref.
\cite{qet}.

Similarly using the rules $(\rho a^\mu + a^\mu \rho)/2 \to \p^\mu
\chi$ and \be \hf(\rho a^\mu a^\nu + a^\mu a^\nu \rho) \to \bigg[
\p^\mu \p^\nu + \hf \Og^{\mu\nu} +{1\over 4} \Og^{\al \mu}\Og^{\bt
\nu}z_\al z_\bt \bigg] \chi, \no \ee the SLD takes \be \mathscr
L_k=\mathscr A^{(k)}_{\mu\nu}(\tilde{a}^\mu \tilde{a}^\nu -
\Sigma^{\mu\nu}) + \mathscr B^{(k)}_{\mu}\tilde{a}^{\mu},
\label{ssld} \ee where in terms of $\mathfrak{M}\equiv \Sigma
\otimes \Sigma+ \Og \otimes \Og/4$, \be \mathscr A^{(k)}_{\mu\nu}
&=& \hf \mathfrak
M^{-1}_{\mu\nu,\al\bt} \p_k\Sigma^{\al\bt}, \label{SA2} \\
\mathscr B^{(k)}_{\mu} &=& \Sigma^{-1}_{\mu\nu} \p_k\lambda^\nu,
\ee and \be \mathscr{F}_{ij}= \hf \mathfrak M^{-1}_{\al\bt,\mu\nu}
\p_j\Sigma^{\al\bt} \p_i\Sigma^{\mu\nu}+\Sigma^{-1}_{\mu\nu}\p_j
\lambda^\mu\p_i\lambda^\nu. \label{sb} \ee Eqs. (\ref{rrld},
\ref{rb}, \ref{ssld}, \ref{sb}) are the main results of our paper.
For the SLD case, they are identical with the results in Ref.
\cite{am2} obtained by a different method.

Finally, we consider the asymptotic attainability of the SLD bound.
Using the relations $\mathrm{Tr}[\rho [\tilde{a}^\al, \tilde
a^\bt]]=\Omega^{\al\bt}$ and \be \mathrm{Tr}[\rho [\tilde{a}^\al
\tilde a^\bt, \tilde a^\mu \tilde a^\nu]] &=&
\Sigma_+^{\al\mu}\Omega^{\bt\nu}+\Sigma_+^{\al\nu}
\Omega^{\bt\mu}\nn && +\Sigma_+^{\mu\bt}\Omega^{\al\nu}+
\Sigma_+^{\nu\bt}\Omega^{\al\mu},\ee the condition (\ref{asym}) can
be simply put in the form \be 4\mathscr A_{\al\bt}^{(i)}\mathscr
A_{\mu\nu}^{(j)} \Sigma ^{\al\mu} \Omega^{\bt\nu}+\mathscr
B_{\al}^{(i)}\mathscr B_{\bt}^{(j)} \Omega^{\al\bt}=0. \label{asymp}
\ee

\section{Applications}

In this section, we take the Gaussian state as a probe to consider
the estimation of the single phase, the two conjugate parameters
in the displacement operator, the damping and temperature of a
bosonic channel, and the squeezing and phase in the squeezing
operator.

\subsection{Phase estimation with two-mode squeezed vacuum}

First, we consider the lossy quantum optical metrology with a
two-mode ($a_1,a_2$) squeezed vacuum (TMSV) \cite{book} as a
consistent check of the obtained results in Refs.
\cite{tmsv,lossv}. The TMSV is a Gaussian state with the
covariance matrix \cite{book} \be \Sigma_{\rm in} = \hf \bigg(
\begin{array}{cccc}
\sigma_x \cosh 2r & -\mathds 1 \sinh 2r \\
-\mathds 1 \sinh 2r & \sigma_x \cosh 2r \\
\end{array}
\bigg). \label{cov} \ee It is worthy to point out that in Ref.
\cite{lossv}, the TMSV is feeded into the interferometer after the
first beam splitter (BS), whereas in Ref. \cite{tmsv}, it is feeded
before the first BS. To see the effects of loss on the ideal setup,
we calculate the relevant results for both of cases.

For the first case, the input-output relation is given by \be
a_1^{\rm out} &=& e^{\imath \phi}a_1^{\rm
in}\sqrt{\epsilon_1}+\upsilon_1\sqrt{1-\epsilon_1}, \nn a_2^{\rm
out} &=& a_2^{\rm in}\sqrt{\epsilon_2}+
\upsilon_2\sqrt{1-\epsilon_2}, \label{losse} \ee where
$\epsilon_1,\epsilon_2$ represent transmissivity of light beams,
and notations $a_i,\upsilon_i$ are the $i$-th ($i=1,2$) arm's and
its bath mode's annihilation operators, respectively. The average
excitation number of the bath is assumed as
$\lr{\upsilon_i^\dg\upsilon_i}=N$. The final state is then
characterized by the covariance matrix \be \Sigma_\z {out} = \hf
\left(
\begin{array}{cccc}
0 & d_1 & b e^{\imath \phi} & 0 \\
d_1 & 0 & 0 & b e^{-\imath \phi} \\
b e^{\imath \phi} & 0 &  0 & d_2 \\
0 & b e^{-\imath \phi} & d_2 & 0 \\
\end{array}
\right), \ee where $d_i= \epsilon_i \cosh 2 r+(1-\epsilon_i)(2N+1)$
and $b=-{\sqrt{\epsilon_1 \epsilon_2}} \sinh 2r$. Substituting it
into Eq. (\ref{sb}) yields \be \mathscr {F}(\phi) = {2 b^2 \over
1+d_1 d_2-b^2}, \label{phase1} \ee which is independent of $\phi$.
For $\epsilon_1=\epsilon_2=1$, $ \mathscr {F}=\sinh^2 2r$
\cite{lossv}.

\begin{figure}[t!]
\includegraphics[width=0.6\columnwidth]{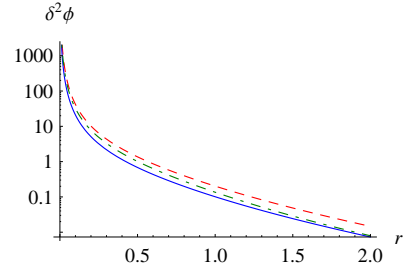}\hspace{0.05in}
\caption{The plots of $\delta^2 \phi$ versus the squeeze parameter
$r$ for the TMSV inserted before (solid) and after
(dashed/dot-dashed) the first BS. Here the average excitation number
$N=0.2$. The transmissivity of light beam $\epsilon_1=0.8$,
$\epsilon_2=0.8 / 1.0$ (dashed/dot-dashed).}
\end{figure}

For the second case, the input-output relation is \be a_1^{\rm
out} &=&e^{\imath \phi}(a_1^{\rm in}+a_2^{\rm in})
\sqrt{\epsilon_1/2}+\upsilon_1\sqrt{1-\epsilon_1}, \nn a_2^{\rm
out} &=&(a_1^{\rm in}-a_2^{\rm in}) \sqrt{\epsilon_2/2}+
\upsilon_2\sqrt{1-\epsilon_2}. \ee The covariance matrix for the
final state is \be \Sigma_\z {out} = \hf \left(
\begin{array}{cccc}
 b_1 e^{2\imath \phi} & d_1 & 0 & 0 \\
 d_1 & b_1 e^{-2\imath \phi} & 0 & 0 \\
0 & 0 & b_2 & d_2 \\
 0 & 0 & d_2 & b_2 \\
\end{array}
\right), \ee where $d_i=\epsilon_i \cosh 2 r+(1-\epsilon_i)(2N+1)$
and $b_i=-{\epsilon_i} \sinh 2r$. We can see that, after the first
BS, the TMSV becomes disentangled, i.e. \be S_2(r) \equiv e^{r(a_1
a_2-a_1^\dg a_2^\dg)} \to e^{r(a_1^2-a_1^{\dg 2})/2}
e^{r(a_2^2-a_2^{\dg 2})/2}, \ee where $S_2(r)$ is the two-mode
squeezing operator. The resulting QFI is given by \be \mathscr
{F}(\phi) ={4 b_1^2 \over 1+d_1^2-b_1^2}, \label{phase2}  \ee which
becomes $ \mathscr {F}=2\sinh^2 2r$ for $\epsilon_1=\epsilon_2=1$
\cite{tmsv}, twice of that for the first case. For the lossy
interferometer, Fig. 1 shows that the quantum entangled state does
not always perform better than coherent but disentangled state for
phase estimation.

Now let us prove the attainability of the two bounds. By inspecting
the structure of $\Sigma_{\rm out}$, we take the measurement scheme
as $M_{\rm after}=\imath(a_1^{\dg 2}-a_1^2)/2$ and $M_{\rm
before}=\imath(a_1^\dg a_2^\dg-a_1a_2)/2$, respectively. The phase
sensitivities are evaluated, \be \delta^2 \phi &=& {\Delta^2 M \over
|{d\lr M/ d\phi}|^2 } \\ &=&
\begin{cases}\dfrac{1+d_1 d_2+b^2(1-2\cos2\phi)}{2 b^2 \cos^2 \phi},
\cr \dfrac{1+d_1^2+b_1^2(1-3\cos 4\phi)/2}{4 b_1^2 \cos^2 2\phi
}.\end{cases} \no \ee The optimization of $\delta^2 \phi$ over
$\phi$ is achieved at $\phi=0$, which are just Eqs. (\ref{phase1})
and (\ref{phase2}).

\subsection{Estimation of two conjugate
parameters in the displacement operator}

Next, we jointly estimate the two conjugate parameters $\ld_R$ and
$\ld_I$ of the displacement operator $D(\ld)=e^{\ld a_1^\dg-\ld^*
a_1}$ with a measurement on the displaced state $\rho=D(\ld)\rho_0
D^\dg(\ld)$ \cite{pomo}. If we take the two-mode squeezed thermal
state $\rho_0=S_2(r) (\rho_{\nu_T} \otimes \rho_{\nu_T})
S_2^\dg(r)$ as the input, where \be
\rho_{\nu_T}=\frac{1}{\nu_T+1}\sum_{n=0}^\infty
\lx\frac{\nu_T}{\nu_T+1}\rx^n \lg n \rg n \ee is a single-mode
thermal state with average excitation number $\nu_T$, the mean
displacement and covariance matrix of this Gaussian state are
given by $\ld_{\rm in}=0$ and \be \Sigma_{\rm in} = {2\nu_T+1
\over 2} \left(
\begin{array}{cccc}
\sigma_x \cosh 2r & -\mathds 1 \sinh 2r \\
-\mathds 1 \sinh 2r & \sigma_x \cosh 2r
\end{array}
\right). \ee

If we include the possible photon loss described by Eqs.
(\ref{losse}) with $\phi=0$ before the displacement operator, the
final state is characterized by $\ld_{\rm out}=(\ld,\ld^*,0,0)^T$,
and \be \Sigma_\z {out} = \hf \left(
\begin{array}{cccc}
d_1 \sigma_x & b \mathds 1 \\
b \mathds 1 & d_2 \sigma_x \\
\end{array}
\right), \ee where $d_i=\epsilon_i (2\nu_T+1)\cosh 2
r+(1-\epsilon_i)(2N+1)$ and $b=-{\sqrt{\epsilon_1 \epsilon_2}}
(2\nu_T+1)\sinh 2r$. The two bounds can be straightforwardly
evaluated from Eqs. (\ref{rbound}), (\ref{sbound}), (\ref{rb}), and
(\ref{sb}), obtaining \be B_R &=& \frac{d_1}{2}+{d_2 b^2 \over
2(1-d_2^2)}+\bigg| \hf+{b^2 \over 2(1-d_2^2)} \bigg|, \nn B_S &=&
{d_1 \over 2}-{b^2 \over 2 d_2}.\ee For lossless case
$\epsilon_1=\epsilon_2=1$, they are in agreement with those in Ref.
\cite{pomo}. Using the same homodyne measurement scheme proposed in
Ref. \cite{pomo}, the sum the two resulting variances is
$B_M=(d_1+d_2)/2-b$.

Fig. 2 displays the three bounds $B_{R,S,M}$ versus the squeezing
parameter $r$. We see that which bound is tighter depends on the
actual values of $r$, and the bound $B_M$ from the homodyne
measurement is always higher than the theoretical RLD and SLD
bounds. We also note that $B_M$ stays much closer to the theoretical
bound for the balanced losses than for the unbalanced losses in the
two arms.

\begin{figure}[t!]
\centering \subfigure[]{ \label{Fig.sub.a}
\includegraphics[width=0.47\columnwidth]{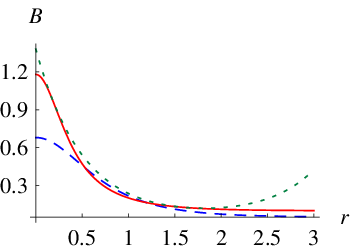}}\hspace{0.05in}
\subfigure[]{ \label{Fig.sub.b}
\includegraphics[width=0.47\columnwidth]{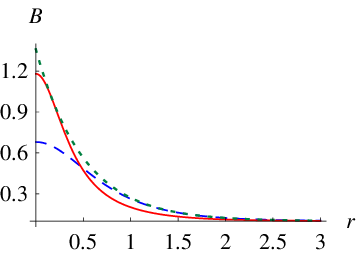}}\hspace{0.05in}
\caption{The plots of $B$ versus the squeeze parameter $r$ for the
estimation of position and momentum. (a) $\epsilon_1=0.9$,
$\epsilon_2=1.0$ for the unbalanced losses in two arms. (b)
$\epsilon_1=\epsilon_2=0.9$ for the balanced losses in two arms.
Here the solid (dashed) lines are for the RLD (SLD) bound, and the
dotted lines represent the bound from the homodyne measurement. The
other parameters are $\nu_T=0.2$ and $N=0$.}
\end{figure}

\subsection{Estimation of damping and temperature}

Then, we consider the problem of estimating the parameters of a
Gaussian channel \cite{am1} describing the evolution of a bosonic
mode $a_1$, coupled with strength $\ga$ to a thermal bath mode
$\upsilon_1$ with mean excitation number $N$. The completely
positive dynamics of the mode $a_1$ in the interaction frame under
the Markovian approximation is represented by the unitary
transformation \be a_1^{\rm out}=a_1^{\rm in} e^{-\ga/2} +
\upsilon_1 \sqrt{1-e^{-\ga}}. \ee We first take the single-mode
state parameterized by \be \rho_{\rm
in}=D(\ld)S(r)\rho_{\nu_T}S^\dg(r)D^\dg(\ld) \label{single} \ee as
the input, where the single-mode squeezing operator
$S(r)=e^{r(a^2-a^{\dg 2})/2}$. The mean displacement and covariance
matrix of this state are given by $\ld_{\rm in}=(\ld,\ld^*)^T$ and
\be
\Sigma_{\rm in}=\frac{2\nu_T+1}{2}\lx \bx -\sinh 2r & \cosh 2r \\
\cosh 2r & -\sinh 2r \ex \rx. \ee The final state would be described
by $\ld_{\rm out}=(e^{-\ga/2}\ld,e^{-\ga/2}\ld^*)^T$ and \be
\Sigma_{\rm out}=\frac{1}{2}\lx \bx b & d \\
d & b \ex \rx, \ee where $b=-e^{-\ga}(2\nu_T+1)\sinh 2r$ and
$d=e^{-\ga}(2\nu_T+1)\cosh 2r+(1-e^{-\ga})(2N+1)$. Since the final
expressions for the QFI matrices are too lengthy, we only display
them numerically in Fig. 3 (a). We see that the SLD bound is tighter
than the RLD bound, as the asymptotic attainability condition
(\ref{asym}) is fulfilled here.

If we take the TMSV ($a_2$ being the ancillary mode) as the input
with the mean displacement $\ld_{\rm
in}=(\ld_1,\ld_1^*,\ld_2,\ld_2^*)$ and covariance matrix
(\ref{cov}), the final state will be described by $\ld_{\rm
out}=(e^{-\ga/2}\ld_1,e^{-\ga/2}\ld_1^*,\ld_2,\ld_2^*)$, and \be
\Sigma_{\rm out} = \hf \bigg(
\begin{array}{cccc}
d_1 \sigma_x & b \mathds 1 \\
b \mathds 1 & d_2 \sigma_x \\
\end{array}
\bigg), \label{coout} \ee where $d_1=e^{-\ga}\cosh
2r+(1-e^{-\ga})(2N+1)$, $b=-e^{-\ga/2}\sinh 2r$, and $d_2=\cosh 2r$.

Noting the form of the covariance matrix and the symmetry of
Eq.(\ref{SA2}), the coefficient matrix $\mathscr A$ of SLD takes the
form \be \mathscr A^{(k)} = \hf \bigg(
\begin{array}{cccc}
d_1^{(k)} \sigma_x & b^{(k)} \mathds 1 \\
b^{(k)} \mathds 1 & d_2^{(k)} \sigma_x \\
\end{array}
\bigg), \quad k=\ga,N. \label{sa2} \ee Actually we do not even know
the explicit expression for the SLD to prove the asymptotically
attainability of the SLD bound in such a case, but recall Eqs.
(\ref{asymp}), (\ref{coout}), (\ref{sa2}) and the fact of $\mathscr
B^{(N)}=0$ implied by $\p_N \ld_{\rm out}=0$. Eqs.(\ref{rb}) and
(\ref{sb}) give the results \be \m F_{\ga \ga}= {1\over \xi^2}+{\xi
t+2 \over 8Y\xi^2}, \quad \m F_{N N} = {1\over Y}, \quad \m F_{\ga
N} = {y \over 2 Y\xi}, \label{rrr} \ee and \be \mathscr F_{\ga \ga}
&=& {2\xi n(n+1)+t- 2
 \over \xi(\xi t+2)}, \label{sss} \\ \mathscr F_{N N}&=&
{\xi t \over Y(\xi t+2)}, \quad \mathscr F_{\ga N} = {2(2n+1) \over
\xi t+2},  \no \ee where $\xi=e^\ga-1$, $n=\sinh^2 r$, $y=2N+1$,
$t=y(2n+1)+1$, and $Y=N(N+1)$. It is verified that Eqs. (\ref{sss})
always give a tighter bound than Eqs. (\ref{rrr}), as indicated by
Fig. 3 (a), in comply with the asymptotic attainability of the SLD
bound. We also note that the TMSV gives better precision than the
single-mode state.

It is also noticed that when the damping is extremely small $\xi \to
0$, no information of temperature is gained, i.e. $\delta N \to
\infty$. On the other hand, when the damping is extremely large $\xi
\to \infty$, or the output state is in equilibrium with the thermal
bath, no information of damping is gained, i.e. $\delta \ga \to
\infty$, and $\delta N \to 1/Y$. By contrast, there might exit some
typos for Eqs. (B9a)-(B9d) in Ref. \cite{am1}, which contain the
erroneous extra factor of $e^{2\ga}$.

\begin{figure}[t!]
\centering \subfigure[]{ \label{Fig.sub.a}
\includegraphics[width=0.47\columnwidth]{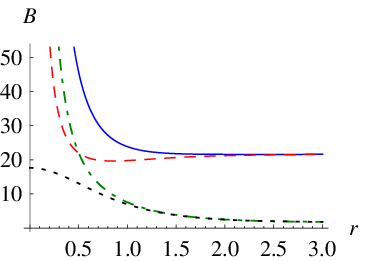}}\hspace{0.05in}
\subfigure[]{ \label{Fig.sub.b}
\includegraphics[width=0.47\columnwidth]{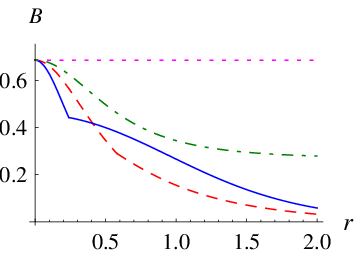}}\hspace{0.05in}
\caption{(a) The plots of $B$ versus the squeeze parameter $r$ for
the estimation of damping and temperature. The solid/dashed
(dot-dashed/dotted) line is for the $B_S/B_R$ with the single-mode
squeezed (TMSV) state. Here $N=0.9$, $\xi=0.5$, $\ld=0$, and
$\nu_T=0$. (b) The plots of $B=\max\{B_R,B_S\}$ versus the probe
energy $n$ (expressed in terms of the effective squeeze parameter
$r$ via $n=(\nu_T+1/2)\cosh 2r-1$) for the estimation of squeezing
and phase. The solid/dot-dashed (dotted) line is for the single-mode
coherent/thermal (squeezed) state. The dashed line is for the
two-mode squeezed thermal state. Here $s=1$, $\varphi=0$, and
$\nu_T=0.1$.}
\end{figure}

\subsection{Estimation of squeezing and phase}

Finally, we address the estimation of $\theta=(s,\varphi)^T$, i.e.
the squeezing and phase parameters in $\zeta=s e^{2 \imath \varphi}$
of the squeezing operator $S(\zeta)=e^{(\zeta^* a_1^2-\zeta a_1^{\dg
2})/2}$ with a measurement on the state $\rho_{\rm
out}=S(\zeta)\rho_{\rm in} S^\dg(\zeta)$. In order to determine the
precision attainable with different Gaussian states, we first take
the single-mode state parameterized by Eq. (\ref{single}) as the
input. The mean displacement and covariance matrix of the final
state can be evaluated by the relations $D^\dg(\ld) a_1
D(\ld)=a_1+\ld$ and $S^\dg(\zeta) a_1 S(\zeta)=a_1 \cosh s-a_1^\dg
e^{\imath \varphi}\sinh s$. From Eqs. (\ref{rb}) and (\ref{sb}), the
RLD and SLD QFI matrices are obtained, \be \m {F}_{\varphi\varphi}
&=& {y^2(2Y+1) \over 2Y^2}\sinh^2 2s+{4y\over Y} |\ld|^2 e^{-2r-2s}
\sinh ^2 s, \nn \m {F}_{ss}&=&{y^2(2Y+1) \over 2Y^2}+ {y \over
Y}|\ld|^2 e^{2r}, \nn \m F_{s\varphi} &=& \imath \lx {y^3 \over 2
Y^2 } \sinh 2s + {2 \over Y} |\ld|^2 e^{-s}\sinh s\rx, \ee and \be
\mathscr{F}_{\varphi\varphi}&=& {2 y^2 \over 2Y+1}\sinh^2 2s+{16
\over y}|\ld|^2 e^{-2r-2s} \sinh ^2 s, \nn \mathscr {F}_{ss}&=&
{2y^2 \over 2Y+1}+ {4 \over y}|\ld|^2 e^{2r}, \quad \mathscr
{F}_{s\varphi}=0, \ee where $y=2\nu_T+1$, $Y=\nu_T(\nu_T+1)$, and
the phase parameter is assumed around $\varphi=0$ for simplicity. In
order to compare the performances for the estimation, we consider
the three different situations: coherent state ($r=0$), squeezed
state ($\ld=0$), and thermal state ($r=\ld=0$). The respective
energies are given by $n=|\ld|^2+\nu_T$, $n=(\nu_T+1/2)\cosh
2r-1/2$, and $n=\nu_T$.

Now we turn to the two-mode squeezed thermal state $\rho_{\rm
in}=S_2(r)(\rho_{\nu_T} \otimes \rho_{\nu_T}) S_2^\dg(r)$ ($a_2$
being the ancillary mode) as the input, and its energy is
$n=(\nu_T+1/2)\cosh 2r-1/2$. The QFI matrices can also be calculated
from Eqs. (). The full expressions for $\m F$ and $\mathscr F$ are
too lengthy to put here, we only display the numerical results in
Fig. 3 (b). It can be seen that for the single-mode states, the
coherent state gives the best performance among others, and the
squeezed state is even worse than the thermal state. The two-mode
squeezed thermal state performs better than the coherent state only
when the probe energy is larger than some actual value. Moreover,
the RLD bounds are more tighter than the SLD bounds when the probe
energies are relatively lower.

\section{Conclusion}

In this paper, we have studied the quantum Cramer-Rao bounds on the
joint multiple-parameter estimation with the Gaussian state as a
probe. We have derived the explicit forms of the right logarithmic
derivative and symmetric logarithmic derivative operators for the
Gaussian state. We have also calculated the corresponding quantum
Fisher information matrices, and found that they can be fully
expressed in terms of the mean displacement and covariance matrix of
the Gaussian state. We have taken some explicit examples to show the
utility of our analytical results.

\begin{acknowledgments}
The authors would like to think Prof. J. P. Dowling for helpful
discussions. This work is supported by NSFC grand No. 11304265 and
the Education Department of Henan Province (No. 12B140013).
\end{acknowledgments}

\end{document}